\documentclass{aa}  
\usepackage{graphicx}
\usepackage{txfonts}
\usepackage{lipsum}
\usepackage{subcaption}         
\usepackage{lscape}             
\usepackage{placeins}           
\usepackage{booktabs}                         
\usepackage[]{natbib}
\usepackage{graphicx,color}
\usepackage{txfonts}
\usepackage[colorlinks=true, citecolor=blue]{hyperref}
\usepackage[table]{xcolor}
\usepackage{multirow} 
\usepackage{adjustbox}
\usepackage{float}
\usepackage{booktabs}
 \usepackage[utf8]{inputenc}
\usepackage[T1]{fontenc}  
\usepackage{dblfloatfix}

\begin{document}

   \title{KMOS Spectroscopic Follow-up of 16 Star Cluster Candidates of the Sagittarius Dwarf Galaxy}
\titlerunning{KMOS kinematics of Sagittarius cluster candidates}
\authorrunning{Obasi et al.}
   \subtitle{}

%
%
%

 \author{
C.O. Obasi\inst{1}\corrauth{casmiroluabuchukwuobasi@gmail.com}
\and J.~G. Fern\'andez-Trincado\inst{2}\corrauth{jose.fernandez@ubo.cl}
\and D. Minniti\inst{1,3}
\and M. G\'omez\inst{1}
\and E.R. Garro\inst{4}
\and R.K. Saito\inst{5}
\and J. Alonso-Garc\'ia\inst{6}
\and V.D. Ivanov\inst{4}
\and J. Corral-Santana\inst{7}
\and A. Luna\inst{7,16}
\and Z. Guo\inst{8,9}
\and R. Kurtev\inst{8}
\and J. Borissova\inst{8}
\and V. Fermiano\inst{8}
\and C. Morris\inst{8}
\and B. Dias\inst{1}
\and C. Caceres\inst{1}
\and S. Saroon\inst{1}
\and J.B. Pullen\inst{1}
\and P.I.A. Rodr\'iguez Cuevas\inst{1}
\and M. Lad\inst{1}
\and S. Federle\inst{1}
\and I. Petralia\inst{1}
\and P.W. Lucas\inst{10}
\and D. Bhadrakumar Sindhu\inst{6}
\and P. Esteves\inst{5}
\and M.G. Navarro\inst{11}
\and T. Palma\inst{12}
\and L. D. Baravalle \inst{12,17}
\and  E. O. Schmidt \inst{12}
\and D. Galdeano\inst{13}
\and M.V. Alonso\inst{12,17}
\and J.L. Nilo Castell\'on\inst{14}
\and A.-N. Chen\'e\inst{15}
\and J. Osses\inst{8}
}

 \institute{
Instituto de Astrof\'isica, Departamento de F\'isica y Astronom\'ia, Facultad de Ciencias Exactas, Universidad Andres Bello, Av. Fern\'andez Concha 700, Santiago, Chile.
\and 
Centro de Investigaci\'on en Astronom\'ia, Universidad Bernardo O’Higgins, Avenida Viel 1497, Santiago, Chile
\and
Vatican Observatory, V00120 Vatican City State, Italy
\and
European Southern Observatory, Karl Schwarzschild Str. 2, 85748 Garching bei M\"unchen, Germany
\and
Departamento de F\'isica, Universidade Federal de Santa Catarina, Trindade 88040-900, Florian'opolis, Brazil
\and
Centro de Astronom\'ia (CITEVA), Universidad de Antofagasta, Av. Angamos 601, Antofagasta, Chile
\and
European Southern Observatory, Casilla 19001, Vitacura, Santiago, Chile
\and
Instituto de F\'isica y Astronom\'ia, Universidad de Valpara\'iso, Av. Gran Breta~na 1111, Playa Ancha, Casilla 5030, Chile
\and
Chinese Academy of Sciences South America Center for Astronomy (CASSACA), National Astronomical Observatories, CAS, Beijing 100101, China
\and
Centre for Astrophysics Research, University of Hertfordshire, College Lane, Hatfield AL10 9AB, UK
\and
Istituto Nazionale di Astrofisica (INAF), Rome Astronomical Observatory, Via Frascati 33, 00078 Monte Porzio Catone, RM, Italy
\and
Observatorio Astron\'omico de C\'ordoba, Universidad Nacional de C\'ordoba, Laprida 854, X5000BGR, C\'ordoba, Argentina
\and
Departamento de Geof\'{i}sica y Astronom\'{i}a, CONICET, Facultad de Ciencias Exactas, F\'{i}sicas y Naturales, Universidad Nacional de San Juan, Av. Ignacio de la Roza 590 (O), J5402DCS, Rivadavia, San Juan, Argentina.
\and
Departamento de Astronom\'ia, Universidad de La Serena. Avenida Ra\'ul Bitran Nachary  1305, La Serena, Chile
\and
NSF NOIRLab, 670 N.  A`ohoku Pl., Hilo, HI 96720, USA
\and
Istituto Nazionale di Astrofisica (INAF), Osservatorio Astronomico di Capodimonte, Salita Moiariello 16, 80131 Napoli, Italy
\and
Instituto de Astronom\'ia Te\'orica y Experimental (IATE-CONICET), Laprida 854, X5000BGR, C'ordoba, Argentina
}
  \date{Received September 30, 20XX}

\abstract
{The Sagittarius (Sgr) dwarf spheroidal galaxy hosts several known globular clusters and numerous additional cluster candidates identified toward its crowded central regions. Establishing the nature of these candidates is difficult because apparent spatial and astrometric structure can arise from the underlying Sgr population.}
{We investigate sixteen Minni stellar-cluster candidates with KMOS spectroscopy to determine whether the observed stars are compatible with Sgr and whether any candidate field shows evidence for a distinct stellar system.}
{We analyse $H$-band KMOS spectra of 341 targets from the VVVX--GalCen Spectroscopic Survey, deriving RVs and empirical metallicities. We compare these with an independent Sgr spectroscopic sample and use VVVX photometry and \emph{Gaia} DR3 astrometry to test for localized stellar excesses relative to local
control fields.}
{Of the 341 KMOS targets, 256 (75\%) have RVs within the adopted Sgr interval, and their metallicities broadly overlap the Sgr comparison population. No field shows statistically compelling kinematic evidence for a distinct system after accounting for the multiple candidates examined. Minni~329 shows the strongest spatial excess, but this is not significant after multiple-testing correction and is not accompanied by significant central concentration.}
{The KMOS targets predominantly trace the Sgr population, but Sgr compatibility alone does not confirm the individual candidates as distinct stellar clusters. None of the sixteen fields can be confirmed as a separate stellar system with the present data, although this does not establish that the original candidates are spurious. Deeper imaging and larger spectroscopic samples are required to determine their physical nature.}

   \maketitle
\nolinenumbers

\section{Introduction}

The Sagittarius (Sgr) dwarf spheroidal galaxy is one of the clearest examples of the ongoing accretion of a satellite by the Milky Way. Since its discovery \citep{ibata1994dwarf}, its main body and extended tidal debris have been traced over a large area of the sky \citep{majewski2003two,newberg2003sagittarius,belokurov2006field,correnti2010northern,vasiliev2021tango}. Its stellar populations and associated globular clusters therefore provide important constraints on the formation and disruption of the Sgr system and its contribution to the Galactic halo.

Several clusters have been securely associated with Sgr, including NGC~6715 (M54), Arp~2, Terzan~7, Terzan~8, Palomar~12, and Whiting~1, together with systems associated with its tidal debris \citep{law2010sagittarius,forbes2010accreted,bellazzini2020globular}. Near-infrared surveys have subsequently enabled searches for fainter stellar systems in the crowded region surrounding the Sgr main body. Using VVV/VVVX photometry together with \emph{Gaia} EDR3 astrometry, \citet{minniti2021eight,minniti2021discovery} examined 41 Minni candidate stellar overdensities projected toward Sgr. Of these, 20 were classified as new Sgr clusters or strong cluster candidates, whereas the remaining 21 were considered inconclusive or unconfirmed. Some of the latter may represent chance fluctuations in the dense Sgr field, although genuine low-mass or spatially extended systems could also be difficult to confirm through conventional spatial-overdensity and colour--magnitude analyses. Confirmation of even a subset of these candidates as distinct stellar systems would substantially increase the known cluster population associated with the Sgr progenitor.

The physical nature of these candidates remains uncertain. Their identification was based on combinations of spatial overdensities, proper motions (PMs), colour--magnitude diagram (CMD) morphology, and, in some cases, RR~Lyrae stars. Subsequent photometric studies characterized several candidates and estimated their ages,
metallicities, and luminosities under the assumption that they represent genuine stellar systems \citep{garro2021physical}. However, \citet{piatti2021globular} showed that the physical reality of these systems must be assessed individually. Using field-star-decontaminated CMDs and spatial stellar distributions, the study found the strongest evidence for Minni~332, with weaker indications for Minni~342, Minni~348, and Minni~349. Minni~341 was identified as a possible young
cluster, whereas Minni~335, Minni~343, and Minni~344 could not be confirmed as stellar aggregates. More generally, the study showed that stellar fluctuations in the composite Sgr field can produce apparent overdensities and CMD features resembling those expected for a stellar cluster. Proper-motion and CMD compatibility with Sgr must therefore be distinguished from evidence that a candidate constitutes a physically distinct stellar system. Nevertheless, the Minni systems continue to enter recent inventories of the Sgr cluster population. For example, \citet{wang2026properties} included the 20 objects identified in the VVVX studies in an updated compilation of Sgr-associated globular clusters, although radial velocities (RVs) and proper motions remain unavailable for the Minni
objects in that compilation.

Spectroscopic information provides an important additional test of the nature of these candidate fields. The APOGEE data available to \citet{piatti2021globular} contained only a small number of bright RGB stars in some of the candidate fields, limiting a systematic spectroscopic assessment of the individual systems. The larger KMOS samples considered here provide an opportunity to characterize the sixteen observed fields systematically using RVs and metallicity estimates. RVs can efficiently separate much of the Milky Way foreground from the Sgr population, while metallicities provide complementary information on the stellar populations sampled along each line of sight. Neither quantity alone, however, establishes the presence of a cluster, because Sgr field stars can share similar kinematic and chemical properties. This distinction is particularly important here because the KMOS targets were drawn from RGB samples previously selected using VVVX and \emph{Gaia} photometry, with preference given to stars associated with the photometric sequences of the published candidates \citep{minniti2021eight,minniti2021discovery,garro2021physical}.

In this work, we present homogeneous KMOS $H$-band spectroscopic follow-up of RGB stars toward sixteen Minni candidate fields observed as part of the VVVX--GalCen Spectroscopic Survey (G\'omez et al., in prep.). We derive individual RVs and metallicity estimates and compare the chemo-kinematic properties of the observed stars with an independent spectroscopic sample of the Sgr field \citep{hasselquist2021apogee}. We additionally use VVVX photometry and \emph{Gaia} DR3 astrometry to test whether the published candidate positions show localized stellar excesses relative to nearby control regions. Our primary aim is to characterize the stellar samples observed toward these candidates and to determine the extent to which the original photometric selection isolates Sgr stars. The present observations are not designed to establish the bound nature of the sixteen candidates; instead, we examine whether any individual fields show spectroscopic or spatial features that warrant further investigation as possible stellar systems.

This paper is organised as follows. Section~\ref{sec:data} describes the KMOS spectroscopy, radial-velocity and metallicity measurements, and the photometric and astrometric data used in this work. Section~\ref{sec:results} presents the spectroscopic comparison with the Sgr population and examines the chemo-kinematic properties of the sixteen candidate fields. Section~\ref{sec:discussion} discusses the combined spectroscopic and spatial evidence for the nature of the Minni candidates, and Section~\ref {sec:conclusions} summarises the main conclusions.

\section{Sample, observations, and spectroscopic analysis}
\label{sec:data}

\subsection{Candidate sample and KMOS target selection}
\label{sec:sample}
In this study, we present a KMOS spectroscopic follow-up of sixteen Minni stellar-cluster candidates projected against the central regions of the Sagittarius dwarf spheroidal galaxy. The candidates were originally identified from VVVX near-infrared photometry and \emph{Gaia} EDR3 astrometry \citep{minniti2021eight,minniti2021discovery}. Their published coordinates and the numbers of stars observed with KMOS are listed in
Table~\ref{table:minni_observations}. The original studies reported 20 candidates. However, VVVX--GalCen observations are currently available for sixteen of them; the remaining four had not yet been observed at the time of this analysis. The present sample therefore reflects the availability of KMOS spectroscopy and represents neither the complete sample nor a statistically selected subset of the published Minni candidates. The KMOS targets were drawn from the proper-motion-decontaminated samples used in the previous photometric studies \citep{minniti2021eight,minniti2021discovery,garro2021physical}, with preference given to stars lying along the identified RGB sequences. The spectroscopy consequently samples photometrically preselected stars rather than providing an unbiased spatial census of each candidate field. Agreement of the observed targets with the previously identified CMD sequences is therefore not treated as independent evidence for the physical nature of the candidates.

\begin{figure*}
    \centering
    \includegraphics[width=0.5\linewidth]
    {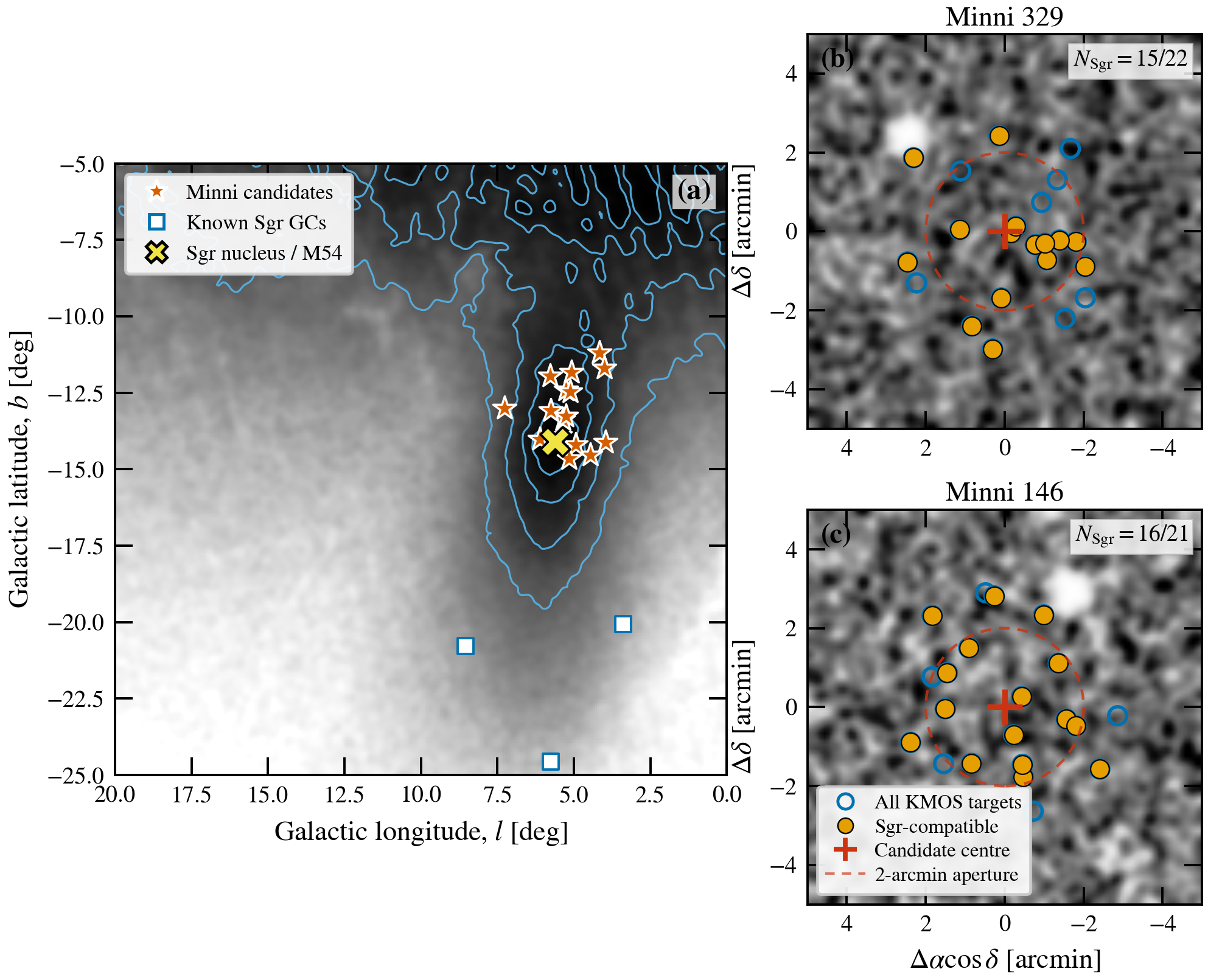}
    \caption{Spatial context of the sixteen Minni candidate fields and representative KMOS target distributions. Panel (a) shows the candidate positions relative to Sgr and its nucleus, M54, with established Sgr globular clusters shown for reference. The grey-scale background in panel (a) shows the smoothed surface-density distribution of Gaia DR3 sources selected with broad Sgr-like proper motions, parallaxes, and magnitudes; the blue contours mark the 70th, 84th, 93rd, and 98th percentiles of the same smoothed density field.} Panels (b) and (c) show $10\times10$ arcmin regions centred on Minni~329 and Minni~146, respectively. Blue circles mark KMOS targets with measured RVs, and orange symbols identify those satisfying the Sgr-compatible velocity criterion. Crosses mark the published candidate centres, and dashed circles show the $2$-arcmin apertures used in the spatial analysis.
    \label{fig:1}
\end{figure*}

Figure~\ref{fig:1} shows the locations of the sixteen observed fields relative to the main body of Sgr and its central region, with M54 shown for reference. The candidate positions lie approximately $0.2$--$3.2^\circ$ from M54 and are projected against the dense stellar population of Sgr. Consequently, stars observed toward these fields can readily share the astrometric and spectroscopic properties of Sgr without necessarily belonging to distinct stellar systems. This
projected contamination is therefore a central consideration throughout our analysis.
\begin{table*}
\centering
\caption{KMOS observations of the sixteen Minni stellar-cluster candidates} considered in this work.
\label{table:minni_observations}
\begin{tabular}{lrrrrr}
\toprule
Candidate &
RA (deg) &
Dec. (deg) &
$N_{\mathrm{obs}}$ &
Observation date &
$t_{\mathrm{exp}}$ (s) \\
\midrule
Minni 144 & 280.0375 & $-30.5739$ & 21 & 2026-03-14 & 1260 \\
Minni 145 & 283.3667 & $-28.5622$ & 20 & 2026-04-12 &  960 \\
Minni 146 & 283.0500 & $-31.9458$ & 21 & 2026-04-12 & 1300 \\
Minni 147 & 280.4833 & $-30.9353$ & 22 & 2026-05-20 & 1260 \\
Minni 148 & 282.9000 & $-30.4161$ & 22 & 2026-05-20 & 1260 \\
Minni 324 & 281.5792 & $-29.4500$ & 20 & 2026-05-20 & 1260 \\
Minni 326 & 282.7792 & $-29.9400$ & 22 & 2026-05-21 & 1260 \\
Minni 328 & 281.8208 & $-30.1000$ & 22 & 2026-05-21 & 1260 \\
Minni 329 & 281.1292 & $-30.0200$ & 22 & 2026-05-21 & 1260 \\
Minni 330 & 283.9083 & $-29.9900$ & 22 & 2026-05-22 & 1260 \\
Minni 335 & 281.8200 & $-30.2600$ & 22 & 2026-05-17 & 1260 \\
Minni 338 & 283.7200 & $-31.6800$ & 19 & 2026-04-13 &  960 \\
Minni 340 & 282.7300 & $-30.4600$ & 21 & 2026-05-22 & 1680 \\
Minni 341 & 283.4900 & $-30.5300$ & 22 & 2026-05-22 & 1800 \\
Minni 342 & 283.5800 & $-31.1200$ & 21 & 2026-05-23 & 1764 \\
Minni 344 & 284.1600 & $-31.1100$ & 22 & 2026-04-12 & 1200 \\
\bottomrule
\end{tabular}

\tablefoot{Candidate coordinates are adopted from \citet{minniti2021eight,minniti2021discovery}. $N_{\mathrm{obs}}$ denotes the number of stars observed with KMOS toward each candidate field. Observation dates and exposure times were verified from the FITS headers of the KMOS data products. The reported
$t_{\mathrm{exp}}$ corresponds to the FITS \texttt{EXPTIME} header value.}
\end{table*}

\subsection{KMOS observations and reduction}
\label{sec:kmos_observations}
The observations were obtained as part of the ESO Public Spectroscopic Survey VVVX--GalCen \citep{arnaboldi2026kmos} (programme ID: 116.29EF.001; PI: M. G\'omez) with the K-band Multi Object Spectrograph (KMOS) on VLT/UT1 at the ESO Paranal Observatory. The KMOS $H$-band grating provides a nominal wavelength coverage of $1.456$--$1.846~\mu{\rm m}$ at a resolving power of $R\simeq4000$. The sixteen candidate fields were observed between March and May 2026, with exposure times of 960--1800~s. The observing
information is summarized in Table~\ref{table:minni_observations}. The data were reduced with the ESO Data Processing System \citep[EDPS;][]{freudling2024adaptive} using the KMOS pipeline (version 4.5.3). The reduction included flat-fielding, illumination and wavelength calibration, and reconstruction of the individual science data cubes. Telluric absorption was corrected using standard-star observations and \textsc{Molecfit} \citep{Smette2015,Kausch2015}, after which the individual exposures were combined.

One-dimensional spectra were extracted from the reduced cubes using a custom Python routine. Source positions were determined from white-light images, and several centroid estimates and circular apertures with radii of 1.0--2.0 spaxels were tested for each target. The centroid--aperture combination maximizing the spectral signal-to-noise ratio (S/N) was adopted. Where a sufficiently populated surrounding annulus was available, a local background estimated from its median
flux per pixel was subtracted from the aperture spectrum. The wavelength intervals 1450--1480~nm and 1800--1868~nm were excluded because of low instrumental sensitivity and strong atmospheric residuals near the edges of the $H$-band setting. Isolated artefacts were rejected using conservative local $7\sigma$ clipping, and no spectral smoothing was applied before the RV measurements. The S/N used for extraction and quality assessment was estimated over 1500--1750~nm from the median flux and a robust noise estimate based on the median absolute deviation.

\subsection{Radial-velocity measurements}
\label{sec:rv_measurements}
RV were measured from the extracted one-dimensional KMOS spectra using \textsc{KROSS} (KMOS Radial Velocities via CrOss-correlation with the VVVX Survey), \citep{Obasi2026KROSS}, a custom Python pipeline developed for the VVVX--GalCen Survey. \textsc{KROSS} cross-correlates continuum-normalized $H$-band spectra with synthetic stellar templates from the PHOENIX--ACES library \citep{husser2013new}. The RV measurements were performed independently of any candidate-specific velocity concentration. Before cross-correlation, the observed spectra were continuum normalized, and wavelength intervals affected by strong telluric residuals, low-sensitivity regions of the spectrum, or identified bad pixels were excluded. The PHOENIX--ACES templates were convolved to the KMOS resolving power ($R\simeq4000$), resampled onto the observed wavelength grid, and restricted to the usable $H$-band wavelength interval.

The template grid used for the final RV measurements spans $3800\leq T_{\rm eff}\leq4800$~K, $1.0\leq\log g\leq3.0$, and $-2.0\leq[\mathrm{Fe/H}]\leq+0.5$, with grid spacings of $\Delta T_{\rm eff}=100$~K, $\Delta\log g=0.5$~dex, and $\Delta[\mathrm{Fe/H}]=0.5$~dex, resulting in 330 synthetic templates. This provides broad coverage of the cool-giant parameter space relevant to the KMOS targets. The grid encompasses the late-type RGB stars targeted by the observations, predominantly K giants, with the coolest templates extending toward the K--M boundary. For each spectrum, the cross-correlation function (CCF) was calculated against the template grid, and the stellar RV was determined from the principal CCF maximum of the best acceptable solution. The atmospheric parameters of the selected template were not interpreted as measurements of the stellar parameters, because the template selection was optimized for the RV determination rather than for atmospheric-parameter inference.

The measured velocities were corrected for the motion of the observer using the target coordinates and observing epoch and are reported in the
barycentric reference frame. For each star, the final RV uncertainty, $\sigma_{\rm RV}$, was obtained by combining the formal uncertainty from
the CCF-peak fit, $\sigma_{\rm fit}$, with the template-to-template dispersion, $\sigma_{\rm temp}$, according to
\begin{equation}
\sigma_{\rm RV} =
\sqrt{\sigma_{\rm fit}^{2}+\sigma_{\rm temp}^{2}}.
\end{equation}

The template contribution was estimated as $1.4826\,{\rm MAD}$ of the RVs obtained from the accepted PHOENIX templates. When this term could not be determined, the formal CCF-fit uncertainty was retained as the reported uncertainty. For the final quality assessment, measurements were classified according
to their total RV uncertainty: A for $\sigma_{\rm RV}<3~{\rm km\,s^{-1}}$, B for $3\leq\sigma_{\rm RV}<6~{\rm km\,s^{-1}}$, and C for $6\leq\sigma_{\rm RV}<15~{\rm km\,s^{-1}}$. 
Measurements with $\sigma_{\rm RV}\geq15~{\rm km\,s^{-1}}$ were assigned lower-quality diagnostic flags but were not removed from the released catalogue.

Finite RV measurements were obtained for all 341 KMOS targets. Of these, 292 measurements are classified as A, 39 as B, and 2 as C, giving 333 measurements, or 97.7\% of the sample, that satisfy the adopted higher-quality criterion. The remaining eight spectra have finite RVs but lower-quality diagnostic flags. These eight measurements are retained in the descriptive field-level statistics and in the released catalogue, with their quality flags reported explicitly. Using the full finite-RV sample, 256 of 341 stars, or 75.1\%, satisfy the adopted Sgr-compatible velocity interval, $120<V_{\rm los}<170~{\rm km\,s^{-1}}$.
The median RV uncertainty of the full sample is $1.57~{\rm km\,s^{-1}}$, with a 16th--84th percentile range of $0.78$--$2.91~{\rm km\,s^{-1}}$. We find only a weak anti-correlation between the RV uncertainty and the spectral S/N (Spearman $\rho=-0.13$, $p=0.016$), indicating that the measurement uncertainty is not determined by spectral S/N alone. The complete catalogue, including the individual velocities, uncertainties,
and quality information, will be made available through CDS. A detailed technical description and validation of \textsc{KROSS} will be presented
by \cite{Obasi2026KROSS}. As part of the KMOS observing programme, we also observe benchmark clusters containing stars with well-established spectroscopic parameters, including $T_{\rm eff}$, $\log g$, $v_{\rm mic}$, $[\mathrm{Fe/H}]$, and $[\alpha/\mathrm{Fe}]$. These observations provide reference samples for validating the spectroscopic measurements derived from the KMOS data. For the present RV analysis, we use the well-studied Galactic globular cluster M22 as an external check of the \textsc{KROSS} velocity zero point. The M22 stars were analysed using exactly the same RV procedure as the Minni targets. For 18
selected M22 stars, we obtained a central velocity of $-146.18~{\rm km\,s^{-1}}$, with a star-to-star dispersion of $2.94~{\rm km\,s^{-1}}$. The corresponding uncertainty on the mean is $0.69~{\rm km\,s^{-1}}$. Compared with the APOGEE determination of $-147.76\pm0.30~{\rm km\,s^{-1}}$ \citep{geisler2025capos}, the
difference is $1.58\pm0.75~{\rm km\,s^{-1}}$, corresponding to approximately $2.1\sigma$. This comparison therefore constrains any global \textsc{KROSS} RV zero-point offset to the level of a few ${\rm km\,s^{-1}}$, well below the velocity differences relevant to the present analysis. We use this comparison only as an external check of the RV zero point; it does not provide an independent calibration of the individual RV uncertainties. 

As a further validation, we compared the PHOENIX-based measurements with velocities obtained from the same KMOS spectra using an independent
cross-correlation implementation following \citet{tonry1979survey} and 27 empirical IRTF templates \citep{rayner2009infrared}. This implementation
is described in detail by \citet{Baravalle2026edr} and \citet{Schmidt2026}. The comparison comprises 97 stars matched by their \emph{Gaia} source
identifiers across the Minni~144, Minni~145, Minni~146, Minni~338, and Minni~344 fields. Defining $\Delta V=V_{\rm PHOENIX}-V_{\rm IRTF}$, we obtain a median offset of $-0.26~{\rm km\,s^{-1}}$ and a robust scatter of $5.70~{\rm km\,s^{-1}}$, where the latter is estimated as 1.4826 times the median absolute deviation. The negligible median offset indicates that the PHOENIX- and IRTF-based measurements are on consistent velocity zero points. We therefore use this comparison as an independent cross-check of the PHOENIX-based RV scale rather than as a validation of the individual RV uncertainties. The resulting RV distributions and their comparison with the Sgr field population are presented in Sect.~\ref{sec:sgr_comparison}.

\subsection{Metallicity estimates}
\label{sec:metallicity}

Stellar metallicities were estimated with the empirical full-spectrum model implemented in \textsc{KROSS-M}, the metallicity extension of the \textsc{KROSS} framework. The procedure has been applied to KMOS spectra of the globular cluster Garro~4 by \cite{Garro2026Surfing}, while a detailed description and validation of the methodology will be presented by \cite{Obasi2026KROSS}. Here, we summarize the aspects relevant to the
present analysis. The model was trained on continuum-normalized APOGEE DR17 spectra and their corresponding ASPCAP $[\mathrm{Fe/H}]$ measurements \citep{garcia2016aspcap,abdurro2022seventeenth}. The parent sample was restricted to high-quality field RGB stars with S/N $\geq120$, $3800\leq T_{\rm eff}\leq5000$~K, $0.5\leq\log g\leq3.2$, and $-2.10\leq[\mathrm{Fe/H}]\leq+0.55$. To obtain broad coverage of this parameter space, stars were sampled jointly in $[\mathrm{Fe/H}]$, $T_{\rm eff}$, and $\log g$, resulting in a final training sample of 294 stars.

The APOGEE spectra were degraded from their native resolving power of $R\simeq22\,500$ \citep{wilson2019apache} to the approximate KMOS resolution of $R\simeq4000$ and resampled onto a common wavelength grid covering $1.5115$--$1.6990~\mu{\rm m}$. The APOGEE and KMOS spectra were processed using the same continuum-normalization procedure. To account for the lower S/N of the KMOS observations, artificial noise realizations at S/N = 15, 25, 40, and 80 were included during model training.

The adopted spectrum-only model uses principal-component analysis (PCA) to compress the normalized spectra, followed by ridge regression \citep{hoerl1970ridge} to
predict the APOGEE ASPCAP $[\mathrm{Fe/H}]$ labels. The final model uses 12 principal components and was evaluated using five-fold cross-validation. Across the 294-star validation sample, the model achieves an RMSE of 0.193~dex, a mean residual of $-0.002$~dex, and a robust residual scatter of 0.162~dex. The residuals show increasing regression toward the centre of the training distribution at its metal-poor end, particularly for $[\mathrm{Fe/H}]\lesssim-1.7$; estimates in this regime are therefore interpreted with additional caution.

As an external validation on native KMOS spectra, the model was applied to independently selected members of M22 (NGC~6656), yielding $[\mathrm{Fe/H}]=-1.67\pm0.02~(\mathrm{stat}) \pm0.19~(\mathrm{cal})$~dex, consistent with the literature metallicity of approximately $-1.70$~dex \citep{harris1996catalogue}. We therefore use the \textsc{KROSS} measurements as empirical metallicity estimates tied to the APOGEE/ASPCAP scale. In the analysis below, these estimates are used to characterize and compare the metallicity distributions of the observed stars rather than as detailed chemical-abundance measurements. Applying this procedure yielded [Fe/H] estimates for all 341 KMOS targets; their distributions and comparison with the Sgr field population are presented in Sect.~\ref{sec:sgr_comparison}.

\subsection{Spatial comparison with local control fields}
\label{sec:spatial_method}
To provide an independent spatial check of the candidate positions, we tested whether localized excesses of Sgr-compatible stars are detectable in the combined VVVX and \emph{Gaia} DR3 catalogues \citep{gaia2023}. Given the crowding and limited depth of the \emph{Gaia}-matched sample,
this analysis is not intended as a definitive test of the underlying stellar density. Rather, it assesses whether the published candidate positions show measurable excesses relative to their local Sgr backgrounds at the depth accessible to the present catalogues. The KMOS targets were not included in this test because they were photometrically preselected and do not constitute spatially complete samples.
We selected stars whose proper motions are statistically consistent with the adopted Sgr proper-motion centre, $(\mu_{\alpha^\ast},\mu_\delta) (-2.75,-1.46)$~mas~yr$^{-1}$ \citep{del2021revealing}. The selection accounts for the individual \emph{Gaia} proper-motion uncertainties, their correlation, and an
intrinsic Sgr dispersion of $0.13$~mas~yr$^{-1}$ added in quadrature \citep{vasiliev2020last}. Stars within the 95\% confidence region of the resulting two-dimensional proper-motion distribution were retained. We additionally applied a deliberately broad near-infrared colour--magnitude selection, $0.45 < J-K_{\rm s} < 1.35$ and $11.0 < K_{\rm s} < 17.5$. The colour interval retains the broadened giant-star population visible in the Sgr fields while reducing contamination from relatively blue foreground stars and very red outliers. The magnitude interval was chosen to avoid the brightest and faintest parts of the catalogue, where saturation at the bright end and crowding, incompleteness, and increasing photometric scatter at the faint end can affect VVV/VVVX photometry \citep{saito2012vvv,surot2019mapping}. These limits were fixed for the full sample and were not intended to isolate the CMD sequence of any individual
candidate. We further required uncertainties smaller than $0.10$~mag in both $J$ and $K_{\rm s}$. Identical criteria were applied to the candidate and control apertures. Selected stars were counted within a circular aperture of radius 2~arcmin centred on each published candidate position. Local comparison
samples were constructed from non-overlapping apertures of the same radius placed at random position angles in two surrounding radial zones, 6--13 and 13--20~arcmin. We attempted to place 25 apertures in each zone. Control apertures were required to lie entirely within the available VVVX footprint, not overlap one another, and remain at least 5~arcmin from any other published candidate position.

\begin{figure*}
    \centering
        \includegraphics[width=0.5\linewidth]
    {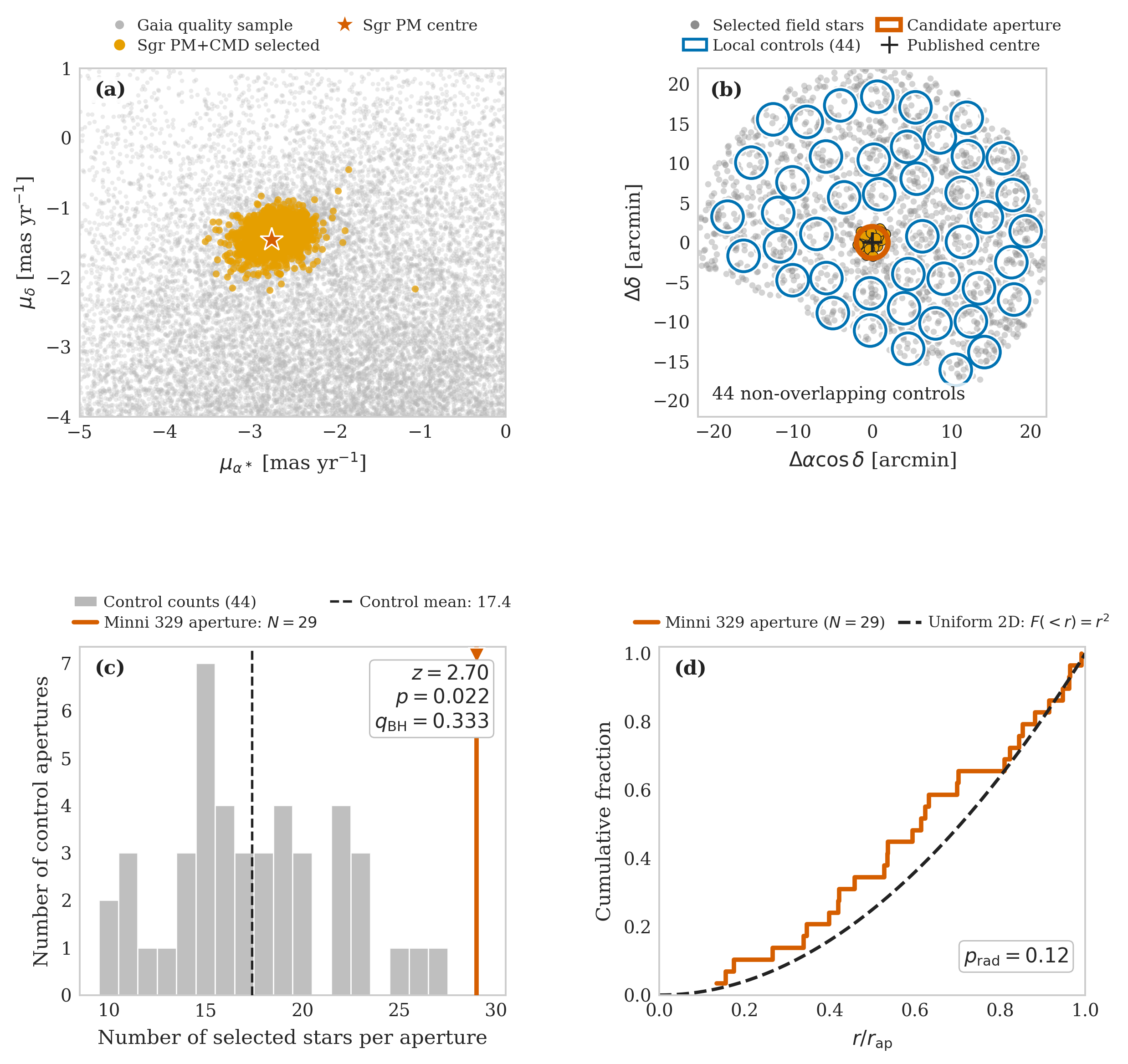}
    \caption{Spatial-control analysis illustrated for Minni~329. Panel (a) shows the proper-motion selection of Sgr-compatible stars. Panel (b) shows the $2$-arcmin candidate aperture and the accepted local control apertures. Panel (c) compares the number of selected stars in the candidate aperture with the distribution of local-control
counts. Panel (d) compares the cumulative radial distribution within the candidate aperture with that expected for a uniform two-dimensional distribution. Minni~329 contains $N=29$ selected stars, with $p_{\rm emp}=0.022$, $q_{\rm BH}=0.333$, and $p_{\rm rad}=0.12$.}
    \label{fig:spatial_control}
\end{figure*}

These criteria yielded 29--49 valid control apertures per candidate. Figure~\ref{fig:spatial_control} illustrates the procedure for Minni~329, which shows the largest positive candidate--control surface-density excess in the sample. Panel (a) shows the proper-motion selection used to identify stars compatible with the adopted Sgr motion. Panel (b) shows the spatial distribution of the selected stars around Minni~329, together with the $2$-arcmin candidate
aperture and the accepted local control apertures. Panel (c) shows the distribution of selected-star counts among the local control apertures, with the count measured in the Minni~329 candidate aperture shown separately by the vertical orange line. Finally, panel (d) compares the cumulative radial distribution of the
selected stars within the candidate aperture with that expected for a uniform two-dimensional distribution.
For each candidate, we calculated the empirical probability of obtaining a control-aperture count at least as large as the candidate count, defined as
\begin{equation}
p_{\rm emp} =
\frac{1+N(n_{\rm control}\geq n_{\rm candidate})}
     {1+N_{\rm control}},
\end{equation}
where $N(n_{\rm control}\geq n_{\rm candidate})$ is the number of control apertures with counts greater than or equal to the candidate count, and $N_{\rm control}$ is the total number of valid control apertures. The resulting probabilities were adjusted across the testable candidate positions using the Benjamini--Hochberg procedure \citep{benjamini1995controlling}, which controls the false-discovery rate when multiple hypotheses are tested. As an additional spatial diagnostic, we examined whether the selected stars within each candidate aperture are preferentially concentrated toward the published candidate centre. Here, $r$ denotes the angular separation of a selected star from the published candidate centre, and $r_{\rm ap}=2$~arcmin is the adopted aperture radius. For a spatially uniform two-dimensional distribution within a circular aperture, the expected cumulative fraction enclosed within radius $r$ is
\begin{equation}
F(<r)=\left(\frac{r}{r_{\rm ap}}\right)^2,
\qquad 0\leq r\leq r_{\rm ap}.
\end{equation}
We compared the observed and expected cumulative radial distributions using a one-sided Kolmogorov--Smirnov test, with small $p$-values indicating a distribution more centrally concentrated toward the published candidate centre than expected for a uniform surface density.

\subsection{Injection--recovery assessment of spatial sensitivity}
\label{sec:injection_recovery}

The absence of a significant spatial excess does not necessarily imply that a stellar system is absent, since the sensitivity of the test depends on both the number of observable cluster stars and their spatial concentration relative to the Sgr background. We therefore performed injection--recovery experiments to quantify the types of stellar systems that would be detectable with the spatial analysis described above. Similar artificial-cluster experiments have been used to characterize the recovery of stellar overdensities in crowded VVVX fields
\citep[e.g.][]{obasi2026consensus}.

Synthetic cluster populations were injected directly into the empirical VVVX--\emph{Gaia} Sgr control fields, thereby preserving the observed stellar density and spatial structure of the local background. The injected stars represent the number of cluster stars, $N_{\rm obs}$, that would already satisfy the astrometric and photometric selection described in Sect.~\ref{sec:spatial_method}; thus, $N_{\rm obs}$ should not be interpreted as the total stellar population or mass of the cluster. Their projected positions were drawn from Plummer profiles \citep{plummer1911problem} with half-light radii $r_h=0.25$, $0.5$, $1$, $2$, and $4$ arcmin. At the adopted Sgr distance of $26.5$ kpc, these correspond approximately to $1.9$, $3.9$, $7.7$, $15.4$, and $30.8$ pc, respectively.

For each combination of $N_{\rm obs}$ and $r_h$, synthetic populations were placed at valid control-field positions within the 16 candidate fields. The same $2$-arcmin aperture and statistical tests used for the real candidates were then applied without modification. A realization was considered jointly recovered when both the aperture-count test ($p_{\rm emp}\leq0.05$) and the radial-concentration test ($p_{\rm rad}\leq0.05$) were satisfied. The resulting recovery fraction therefore measures the sensitivity of the adopted spatial test rather than the intrinsic completeness of the VVVX--\emph{Gaia} catalogue.

\begin{figure*}
    \centering
    \includegraphics[width=0.9\linewidth]{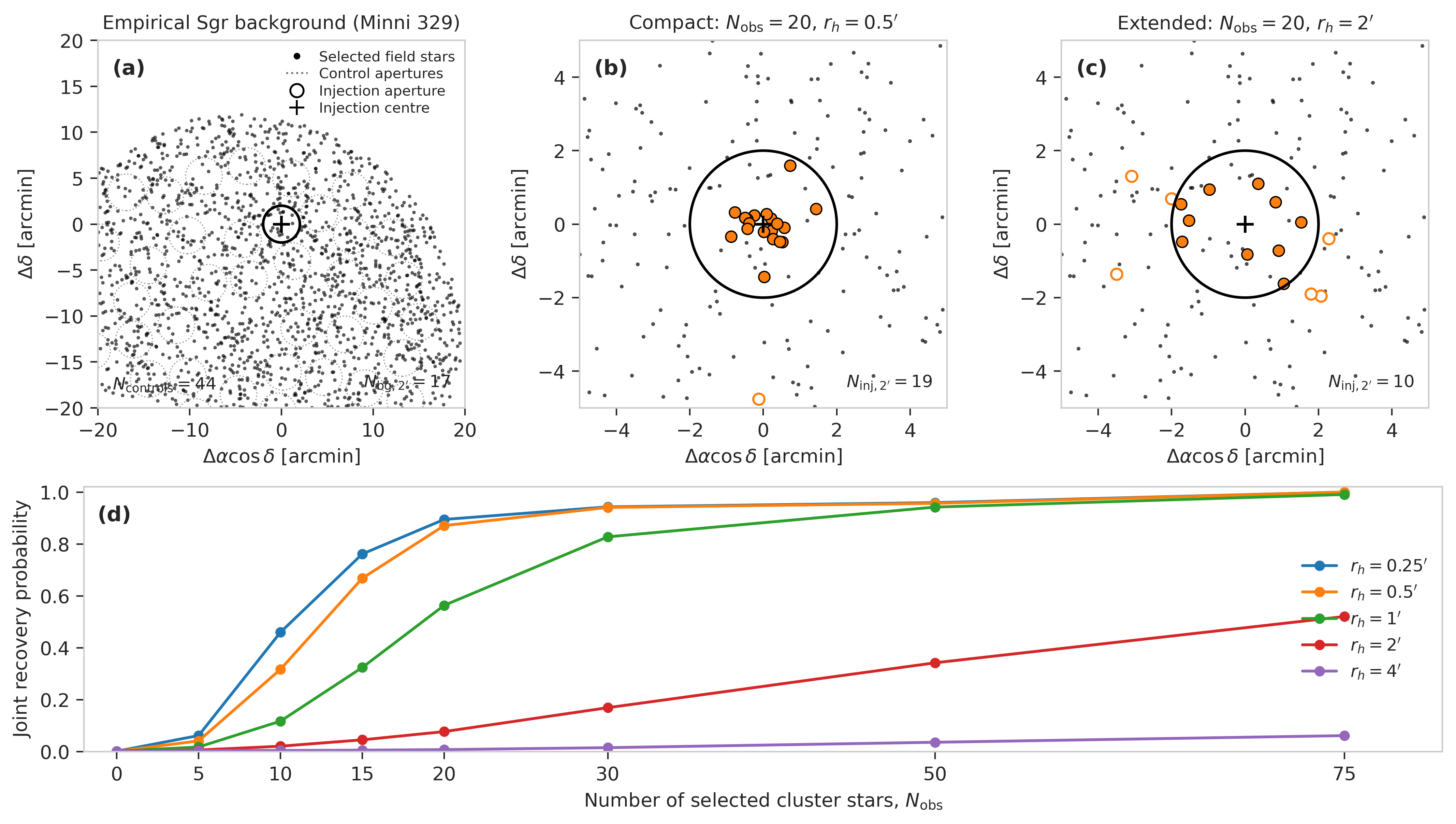}
    \caption{Injection--recovery assessment of the spatial analysis. Panel (a) shows the empirical Sgr background and local control apertures for the representative Minni~329 field. Panels (b) and (c) show illustrative injections of $N_{\rm obs}=20$ stars with $r_h=0.5$ and $2$ arcmin, respectively. Panel (d) shows the joint recovery probability as a function of $N_{\rm obs}$ for the adopted Plummer half-light radii, pooled over the 16 candidate fields. }
\label{injection}
\end{figure*}

Figure~\ref{injection} illustrates the injection--recovery experiment and summarizes its results. Panel (a) shows the empirical Sgr background around a representative control position in the Minni~329 field after applying the same astrometric and photometric selection used in the spatial analysis. The black points represent the selected field stars, the dotted circles show the valid local control apertures, and the solid circle marks the $2$-arcmin aperture used for
the illustrative injections. This representative aperture contains 17 background stars. Panel (b) shows the same local background after injecting $N_{\rm obs}=20$ stars following a compact Plummer profile with $r_h=0.5$ arcmin. In this realization, 19 of the 20 injected stars fall within the adopted $2$-arcmin aperture, producing a clearly concentrated population against the Sgr background. Panel (c) shows an otherwise equivalent realization with a more extended profile, $r_h=2$ arcmin. In this case, only 10 of the 20 injected stars fall within the aperture, illustrating the loss of spatial contrast as the population becomes more extended.

Panel (d) generalizes these illustrative examples by showing the joint recovery probability, pooled over the 16 candidate fields, as a function of $N_{\rm obs}$ for the different adopted half-light radii. Compact populations are recovered efficiently once a sufficient number of selected stars is present. For $N_{\rm obs}=20$, populations with $r_h\lesssim0.5$ arcmin are recovered in approximately 90\% of the realizations. The recovery probability decreases to approximately 60\% at $r_h=1$ arcmin and below 10\% at $r_h=2$ arcmin. Increasing $N_{\rm obs}$ substantially improves the recovery of compact and moderately extended populations, whereas even relatively rich populations with $r_h=4$ arcmin remain difficult to recover. The
comparison therefore demonstrates that the sensitivity of the spatial test depends strongly on both the observable richness and spatial concentration of the injected population.
We additionally verified the calibration of the radial-concentration statistic using both idealized uniform realizations and the empirical Sgr control fields. For $10^5$ uniform realizations, the fraction satisfying $p_{\rm rad}\leq0.05$ remains within $0.0495$--$0.0518$ for samples containing 5--75 stars. Among the 676 real control apertures, the corresponding fraction is $0.053$. The adopted threshold therefore shows the expected null behaviour against the empirical Sgr background.

\section{Results}
\label{sec:results}
\subsection{Spectroscopic properties relative to the Sagittarius field}
\label{sec:sgr_comparison}

Reliable RVs were obtained for all 341 KMOS targets observed across the sixteen candidate fields. The measured velocities span a broad range, reflecting contributions from both the Sgr population and foreground/background stars along these crowded lines of sight. To characterize the Sgr component independently of the Minni candidates, we compared the KMOS measurements with the spectroscopic Sgr sample of \citep{hasselquist2021apogee}.

\begin{figure*}
    \centering
    \includegraphics[width=0.9\linewidth]
    {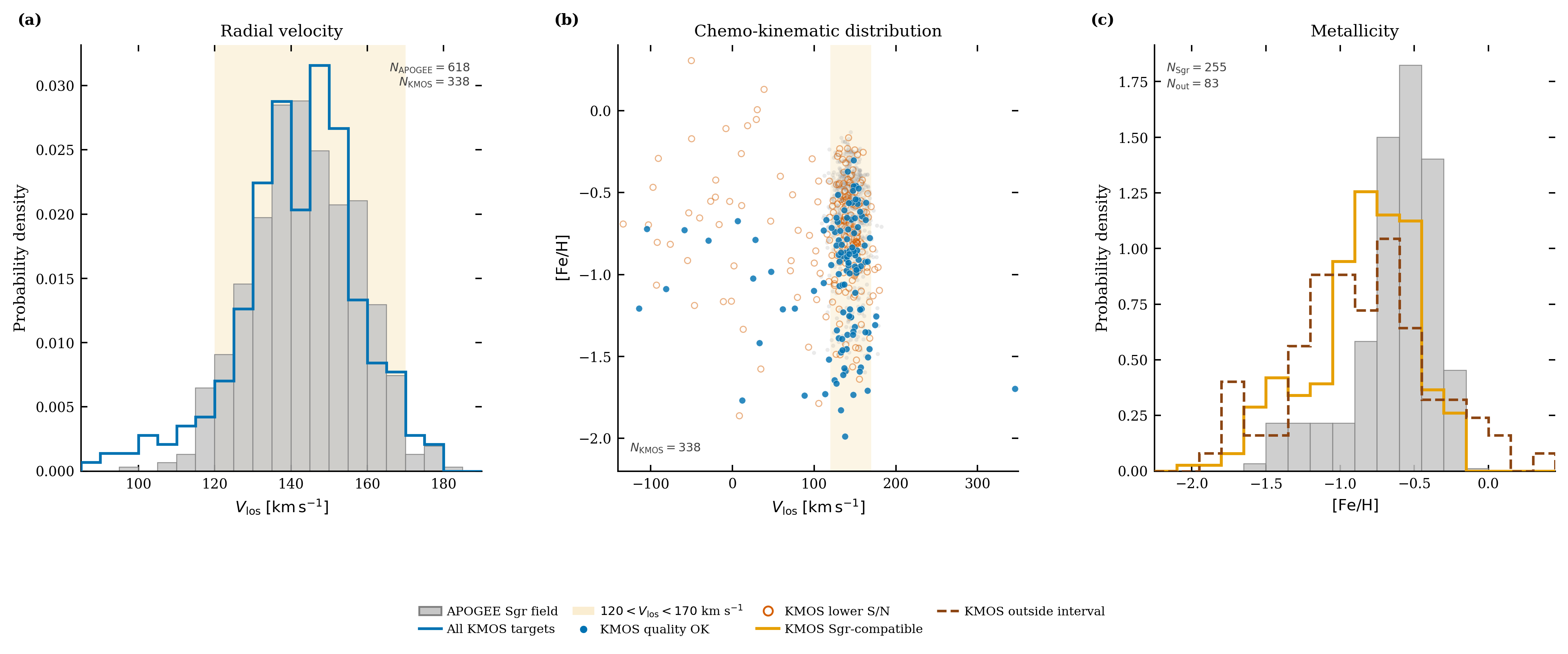}
    \caption{Chemo-kinematic comparison of the KMOS targets with the independent APOGEE Sgr field sample. Panel (a) shows the normalized line-of-sight velocity distributions, with the adopted Sgr-compatible interval ($120<V_{\rm los}<170~{\rm km\,s^{-1}}$) shaded. Panel (b) shows metallicity as a function of line-of-sight velocity; filled blue and open orange symbols denote KMOS measurements satisfying and not satisfying the adopted metallicity-quality criterion, respectively, and grey symbols show the APOGEE comparison sample. Panel (c) shows the normalized metallicity distributions of KMOS targets inside the adopted Sgr velocity interval (orange solid line) and outside this interval (brown dashed line), together with the APOGEE Sgr field distribution in grey.}
    \label{fig:sgr_chemo_kinematics}
\end{figure*}

Figure~\ref{fig:sgr_chemo_kinematics} compares the KMOS chemo-kinematic distributions with the independent Sgr reference sample. The velocity distribution in panel (a) shows the dominant central-Sgr component near $V_{\rm los}\simeq140~{\rm km\,s^{-1}}$. Guided by this distribution, we adopt a single fixed interval,
$120<V_{\rm los}<170~{\rm km\,s^{-1}}$, to identify KMOS targets whose velocities are compatible with the dominant Sgr population. This criterion is applied uniformly to all sixteen fields and is not fitted to the velocity distribution of any individual candidate.

In total, 256 of the 341 KMOS targets (75 \%) have velocities within the adopted Sgr interval. This concentration is also apparent on a field-by-field basis, with
most candidate fields containing a prominent component within this velocity range (Appendix~\ref{app:rv_distributions}; Figure~\ref{fig:rv_app1}). Panel (b) compares the KMOS metallicities and velocities directly with the APOGEE Sgr sample \citep{hasselquist2021apogee}. Metallicities are available for all 341 KMOS targets. Of these, 123 have ${\rm S/N}\geq25$ and are classified as high-quality metallicity measurements; these are shown as filled blue symbols. The remaining 218 measurements have ${\rm S/N}<25$ and are shown as open orange symbols. Panel (c) compares the metallicity distributions of KMOS targets inside and outside the adopted Sgr velocity interval with that of the APOGEE comparison sample. All finite KMOS metallicity measurements are included in this comparison, irrespective of their quality classification. The metallicities of the targets within the Sgr velocity interval overlap the broad distribution of the APOGEE Sgr
sample, although the KMOS targets preferentially sample its relatively metal-rich component. This behaviour is consistent with the RGB-based target selection and with the metallicities expected from previous photometric analyses \citep{garro2021physical}.

The RV and metallicity distributions show that the photometrically preselected KMOS targets predominantly sample stars that are spectroscopically compatible with the Sgr population. This classification is used only to isolate the dominant Sgr-compatible spectroscopic component and is not interpreted as cluster membership, because stars in the surrounding Sgr population naturally occupy the same velocity and metallicity space. The field-to-field properties of these stars are examined in Sect.~\ref{sec:field_results}.

\subsection{Field-by-field spectroscopic properties}
\label{sec:field_results}

The contribution of Sgr-compatible stars varies among the sixteen candidate fields. Table~\ref{tab:field_results} summarizes the field-level spectroscopic measurements using the fixed velocity criterion defined in Sect.~\ref{sec:sgr_comparison}. The fraction of Sgr-compatible targets ranges from 0.50 in Minni~147 to 0.91 in Minni~330, Minni~335, and Minni~344, with more than half of the observed stars satisfying the criterion in the sixteen fields. For each field, we report the median velocity and robust velocity scatter of this subset, together with the corresponding median metallicity and robust metallicity scatter. These quantities characterize the populations sampled along each line of sight and are not interpreted as cluster systemic velocities, intrinsic velocity dispersions, or cluster metallicities.

\begin{figure*}
    \centering
    \includegraphics[width=0.7\linewidth]
    {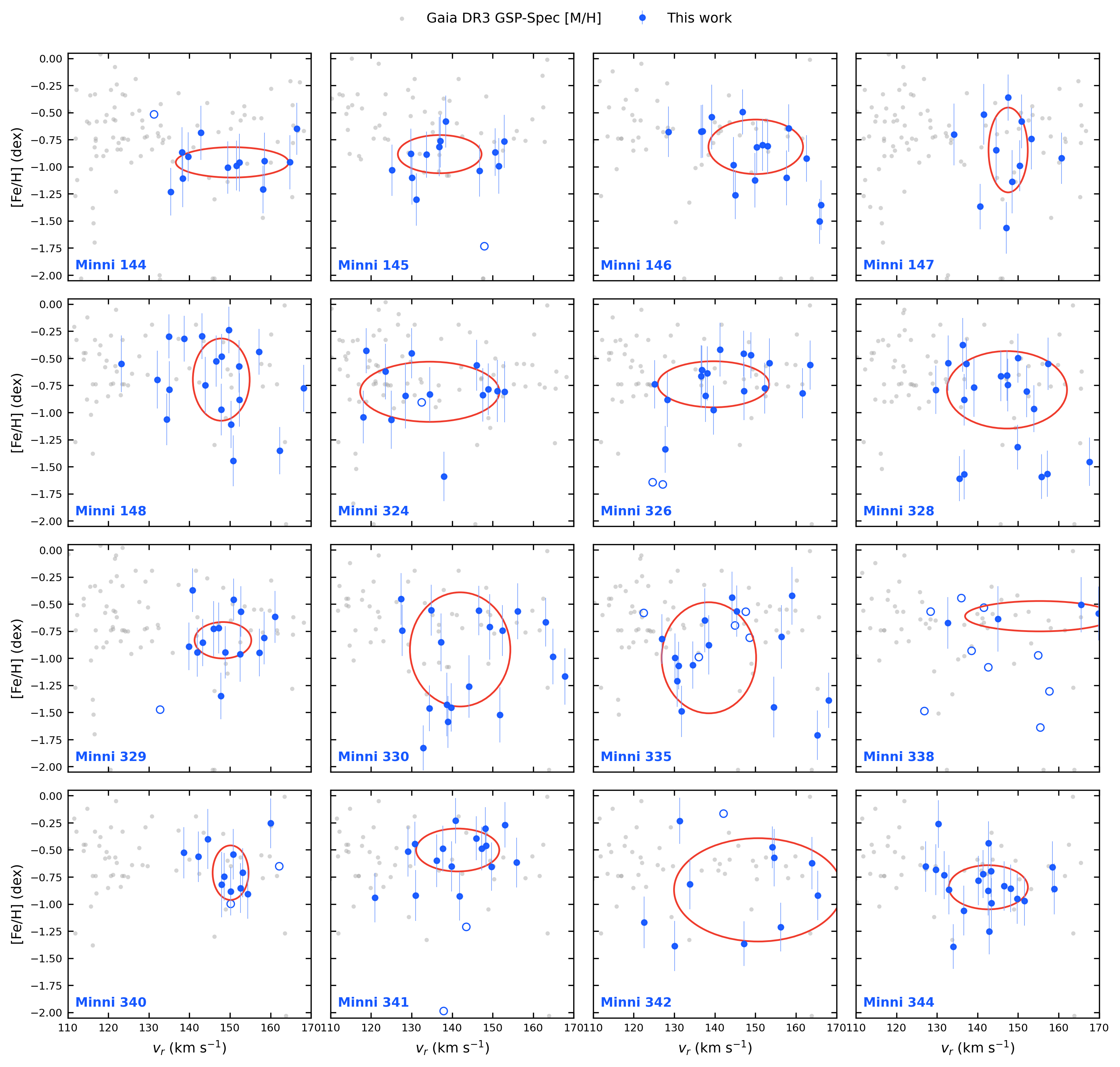}
    \caption{Joint radial-velocity--metallicity distributions for the KMOS targets in the sixteen Minni candidate fields. Filled blue circles show KMOS stars retained in the calculation of the central values for each field, with vertical bars indicating their $1\sigma$ [Fe/H] uncertainties. Open blue circles show KMOS targets excluded from this calculation. Grey points show \emph{Gaia} DR3 sources within $2^\circ$ of each candidate with RVS radial velocities and GSP-Spec [M/H] measurements that satisfy the adopted quality cuts. The red ellipse indicates the robust, MAD-based scatter of the retained KMOS sample in radial velocity and metallicity. The ellipse is intended only as a visual guide and does not represent a formal confidence region. The panels illustrate the differing degrees of apparent chemo-kinematic concentration among the candidate fields. These distributions are shown for comparison and were not used as an additional membership criterion.}
    \label{fig:field_rv_feh}
\end{figure*}

\begin{table*}
\centering
\caption{Field-level spectroscopic properties of the sixteen Minni candidate fields.}
\label{tab:field_results}
\begin{tabular}{lrrrrrrr}
\toprule
Candidate &
$N_{\rm RV}$ &
$N_{\rm Sgr}$ &
$f_{\rm Sgr}$ &
$V_{\rm los}$ &
$\sigma_{V,\rm rob}$ &
$[\mathrm{Fe/H}]$ &
$\sigma_{[\mathrm{Fe/H}],\rm rob}$ \\
&
&
&
&
(km\,s$^{-1}$) &
(km\,s$^{-1}$) &
(dex) &
(dex) \\
\midrule
Minni 144 & 21 & 13 & 0.62 & 149.5 & 14.5 & $-0.96$ & 0.14 \\
Minni 145 & 20 & 13 & 0.65 & 137.1 & 10.7 & $-0.89$ & 0.19 \\
Minni 146 & 21 & 16 & 0.76 & 150.1 & 11.7 & $-0.82$ & 0.25 \\
Minni 147 & 22 & 11 & 0.50 & 147.5 &  4.8 & $-0.85$ & 0.39 \\
Minni 148 & 22 & 19 & 0.86 & 147.9 &  7.0 & $-0.70$ & 0.38 \\
Minni 324 & 20 & 12 & 0.60 & 136.3 & 15.6 & $-0.82$ & 0.09 \\
Minni 326 & 22 & 18 & 0.82 & 138.9 & 15.3 & $-0.76$ & 0.26 \\
Minni 328 & 22 & 19 & 0.86 & 147.3 & 14.8 & $-0.79$ & 0.36 \\
Minni 329 & 22 & 15 & 0.68 & 147.8 &  7.3 & $-0.85$ & 0.19 \\
Minni 330 & 22 & 20 & 0.91 & 139.4 & 16.0 & $-0.87$ & 0.45 \\
Minni 335 & 22 & 20 & 0.91 & 141.4 & 14.8 & $-0.85$ & 0.36 \\
Minni 338 & 19 & 13 & 0.68 & 142.7 & 18.2 & $-0.67$ & 0.34 \\
Minni 340 & 21 & 17 & 0.81 & 148.6 &  8.6 & $-0.71$ & 0.27 \\
Minni 341 & 22 & 18 & 0.82 & 141.4 &  9.5 & $-0.56$ & 0.21 \\
Minni 342 & 21 & 11 & 0.52 & 147.2 & 19.8 & $-0.82$ & 0.52 \\
Minni 344 & 22 & 20 & 0.91 & 142.7 &  9.8 & $-0.85$ & 0.20 \\
\bottomrule
\end{tabular}

\tablefoot{The quantities are calculated for stars satisfying the fixed Sgr velocity criterion, $120 < V_{\rm los} < 170~{\rm km\,s^{-1}}$. They characterise the observed Sgr-compatible samples and should not be interpreted as confirmed cluster systemic velocities, intrinsic dispersions, or cluster metallicities. $N_{\rm RV}$ is the number of targets with reliable radial-velocity measurements, $N_{\rm Sgr}$ is the number satisfying the adopted Sgr velocity criterion, and $f_{\rm Sgr}=N_{\rm Sgr}/N_{\rm RV}$. The listed $V_{\rm los}$ and $[\mathrm{Fe/H}]$ values are the median velocity and median metallicity, respectively, of the Sgr-compatible subset. The quantities $\sigma_{V,\rm rob}$ and $\sigma_{[\mathrm{Fe/H}],\rm rob}$ are the corresponding robust scatters, defined as $1.4826$ times the median absolute deviation. Lower-quality metallicity measurements are retained and should be interpreted together with their individual uncertainties.}
\end{table*}
The individual RV distributions show concentrations within the adopted Sgr velocity range in most fields (Appendix~\ref{app:rv_distributions}; Figure~\ref{fig:rv_app1}), although their widths vary substantially. The robust velocity scatters of the Sgr-compatible subsets range from $4.8~{\rm km\,s^{-1}}$ for Minni~147 to $19.8~{\rm km\,s^{-1}}$ for Minni~342 (see Table~\ref{tab:field_results}).
The joint velocity--metallicity distributions provide a complementary view of this field-to-field variation. Figure~\ref{fig:field_rv_feh} compares the KMOS measurements in each candidate field with an independent Sgr comparison population drawn from \emph{Gaia} DR3 \citep{gaia2023}. Some fields contain several stars occupying a comparatively compact region in both $V_{\rm los}$ and $[\mathrm{Fe/H}]$, whereas others are more broadly distributed and visually difficult to distinguish from the Sgr field. Several fields show intermediate behaviour for which the present spectroscopic sample does not permit a clear distinction.

The apparent concentrations in Figure~\ref{fig:field_rv_feh} are descriptive rather than evidence by themselves for distinct stellar systems. The number of KMOS targets per field is small, the individual metallicity measurements have non-negligible uncertainties, and the Sgr population itself occupies the same broad chemo-kinematic parameter space. In particular, visual concentration in the joint $V_{\rm los}$--$[\mathrm{Fe/H}]$ plane need not correspond to an unusually narrow velocity distribution. Independently of the joint chemo-kinematic appearance, we therefore tested whether any individual field has an RV distribution narrower than expected from sampling the Sgr comparison population. For each candidate, random samples containing the same number of Sgr-compatible stars were drawn from the Sgr velocity distribution, and their robust velocity scatters were compared with the observed value. Minni~147 and Minni~148 give the smallest nominal probabilities, $p_{\rm narrow}=0.0220$ and $0.0228$, respectively. Neither provides compelling evidence once the sixteen candidate fields examined are taken into account. Thus, although these two fields have the most unusual RV widths in the present sample, the velocity distributions alone do not provide statistically compelling evidence for a kinematically distinct component.

\begin{figure}
    \centering
    \includegraphics[width=\columnwidth]
    {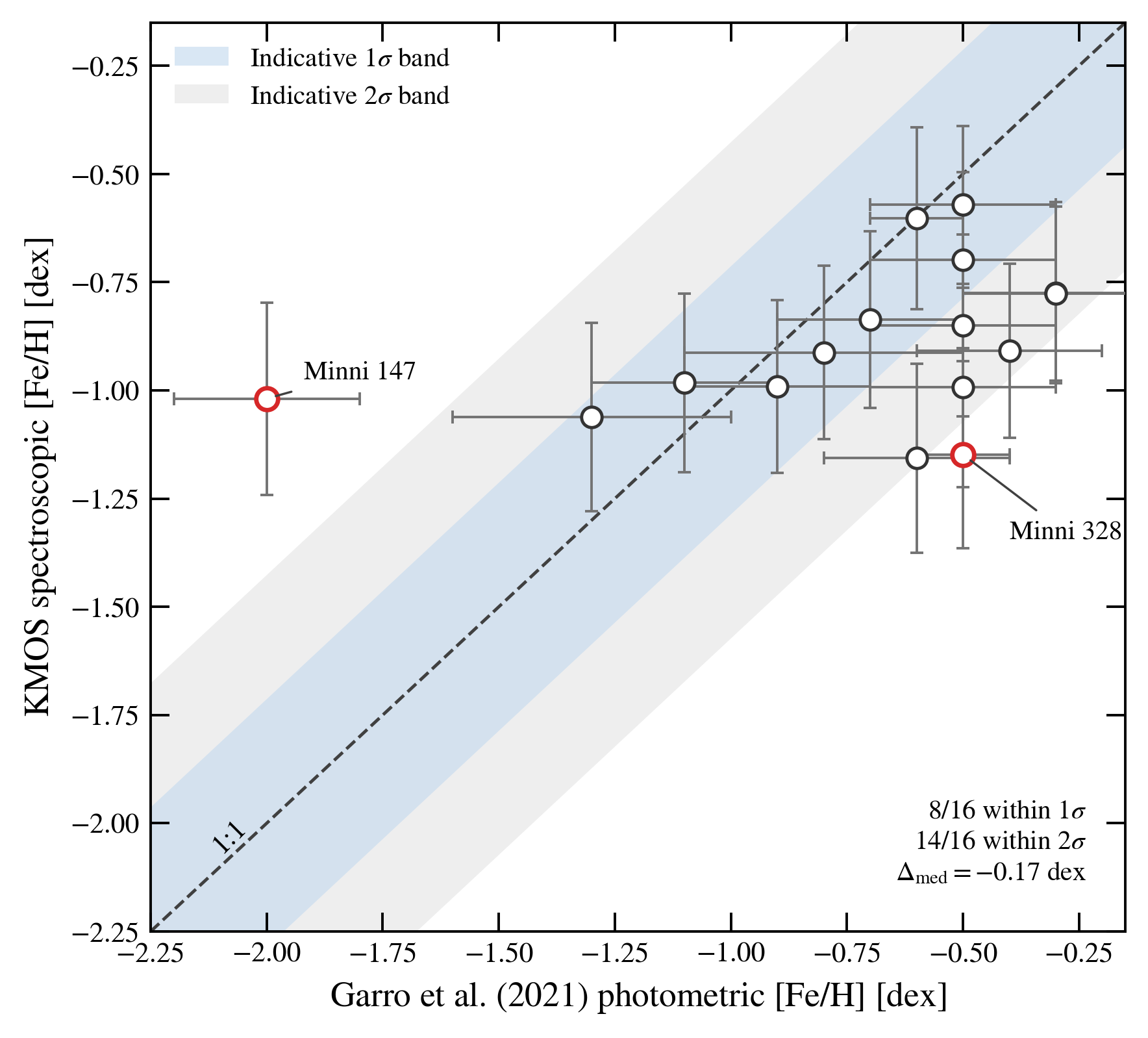}
    \caption{Comparison between the KMOS spectroscopic metallicity estimates and the photometric metallicities reported by  \citet{garro2021physical} for the sixteen candidate fields. Error bars show the uncertainties associated with the two estimates, and the dashed line marks equality. The shaded regions indicate representative $1\sigma$ and $2\sigma$ agreement intervals based on the median combined uncertainty; individual agreement levels are calculated using the combined uncertainty of each field. Minni~147 and Minni~328, the only fields differing by more than  $2\sigma$, are labelled.}
    \label{fig:feh_comparison}
\end{figure}

The spectroscopic metallicity estimates provide an additional comparison with the photometric values reported by \citet{garro2021physical}. Figure~\ref{fig:feh_comparison} compares the two measurements for all sixteen fields. Eight candidates agree within their combined $1\sigma$ uncertainties and fourteen within $2\sigma$; only Minni~147 and Minni~328 differ by more than $2\sigma$. The median difference, $\Delta[\mathrm{Fe/H}]=[\mathrm{Fe/H}]_{\rm KMOS}
-[\mathrm{Fe/H}]_{\rm phot}$, is $-0.17$~dex, indicating a modest tendency for the KMOS estimates to be more metal-poor.

The values reported by \citet{garro2021physical} were inferred from photometric isochrone fitting, whereas the KMOS values characterize the Sgr-compatible stars observed spectroscopically in each field. Their broad agreement therefore indicates that the spectroscopy generally samples stars with metallicities consistent with the populations inferred photometrically, but does not provide independent evidence that the candidates are distinct stellar systems.
\subsection{Spatial distribution of the Sgr-compatible populations}
\label{sec:controls}

We next examined whether the published candidate positions coincide with localized enhancements in the VVVX--\emph{Gaia} Sgr-compatible population using the candidate--control comparison described in Sect.~\ref{sec:spatial_method}.  Overall, the candidate positions do not show a systematic enhancement relative to their local control apertures. The strongest positive signal is found toward Minni~329, whose $2$-arcmin candidate aperture contains 29 selected stars compared with a control-field mean of $17.39\pm4.31$ stars per aperture, where the quoted uncertainty is the standard deviation among the 44 accepted control apertures. The corresponding excess is 11.61 stars, or a candidate-to-control ratio of 1.67. The candidate count lies 2.70 control-field standard deviations above the mean ($z_{\rm emp}=2.70$), and none of the 44 control apertures contains as many selected stars. Under the adopted finite-sample prescription, this gives an empirical probability of $p_{\rm emp}=1/45=0.022$. The nominal Minni~329 excess is not significant when the full set of candidate positions is considered. In addition, the radial distribution of the 29 selected stars within the Minni~329 aperture is not significantly more centrally concentrated than expected for a uniform two-dimensional distribution ($p_{\rm rad}=0.12$; Figure~\ref{fig:spatial_control}d). None of the tested candidates therefore satisfies both a multiple-testing-corrected surface-density criterion and an independent central-concentration criterion.

The spatial analysis consequently provides no statistically compelling evidence for a localized Sgr-compatible stellar component at any of the tested candidate positions. This non-detection should be interpreted within the limitations of the available VVVX--\emph{Gaia} sample, whose effective depth and completeness are restricted by the photometric and astrometric requirements and by crowding in these dense fields. The absence of a significant excess therefore does not establish that the original candidates are spurious; rather, they cannot be independently distinguished from fluctuations in the surrounding Sgr population with the present data. We therefore retain the term \emph{candidate field} and do not interpret the spectroscopically Sgr-compatible KMOS stars as confirmed cluster members.
\section{Discussion}
\label{sec:discussion}
\subsection{What the KMOS spectroscopy establishes}
\label{sec:discussion_spectroscopy}

The principal result of the KMOS follow-up is the presence of a prominent Sgr-compatible spectroscopic component across the sixteen candidate fields. Of the 341 targets with reliable RVs, 256 (75 \%) lie within the independently defined Sgr velocity interval, and the field-by-field distributions generally show concentrations in the same velocity range (Appendix~\ref{app:rv_distributions}; Figure~\ref{fig:rv_app1}). The empirical metallicity estimates of these stars also overlap the broad metallicity distribution of the independent Sgr comparison sample. Taken together, these results show that the photometrically preselected KMOS targets predominantly sample stellar populations that are chemo-kinematically compatible with Sgr.

This result is more limited than establishing the existence of sixteen distinct stellar clusters. Because the candidate fields are projected against the dense central regions of Sgr, ordinary Sgr field stars naturally occupy similar proper-motion, radial-velocity, and metallicity space. The observed field-by-field RV concentrations therefore establish coherence with the Sgr population, but do not by themselves demonstrate kinematically distinct systems. This interpretation is supported by the comparison with the independent Sgr velocity distribution: although Minni~147 and Minni~148 show the narrowest RV distributions, neither remains statistically significant after accounting for the sixteen fields examined.

The comparison with the photometric metallicities of \citet{garro2021physical} provides an additional consistency check. Fourteen of the sixteen fields agree with the earlier estimates within their combined $2\sigma$ uncertainties, with a median difference of $-0.17$~dex in the sense $[\mathrm{Fe/H}]_{\rm KMOS}-[\mathrm{Fe/H}]_{\rm phot}$. Given the different nature of the two measurements, this broad agreement indicates that the KMOS spectroscopy generally samples stellar populations consistent in metallicity with those inferred from the original photometric analysis. The discrepancies for Minni~147 and Minni~328 illustrate, however, that this correspondence is not uniform across the sample.

\subsection{Do any individual candidate fields stand out?}
\label{sec:discussion_candidates}

Although no candidate satisfies the adopted statistical criteria for a distinct stellar system, several fields show features that may warrant further investigation. Minni~147 and Minni~148 have the narrowest Sgr-compatible RV distributions in the sample, with nominal probabilities of approximately 0.02 relative to random draws from the Sgr comparison population. These signals do not remain significant after accounting for the sixteen fields examined, and therefore cannot
be regarded as evidence for kinematically cold systems on their own. Moreover, the joint $V_{\rm los}$--$[\mathrm{Fe/H}]$ distributions show that a narrow RV distribution does not necessarily correspond to strong chemical coherence. More generally, the candidate fields display a range of apparent chemo-kinematic structure, from relatively compact groupings to distributions that are visually difficult to distinguish from the Sgr field.

Minni~329 provides a different type of indication. Its published position contains the strongest surface-density excess in the independent VVVX--\emph{Gaia} analysis, with 29 selected stars compared with a local-control mean of $17.39\pm4.31$. Only one of the 44 control apertures contains at least as many selected stars, corresponding to $p_{\rm emp}=0.022$. However, this nominal excess does not remain significant after accounting for the 16 candidate positions tested and is not accompanied by significant central concentration ($p_{\rm rad}=0.12$). Minni~329 therefore shows the strongest spatial indication in the present analysis, but the available evidence is insufficient to establish a distinct stellar system.

The injection--recovery experiments provide important context for interpreting this and the other spatial non-detections. The adopted analysis is most sensitive to systems that retain a sufficiently populated and compact central component. For example, at $N_{\rm obs}=20$, compact populations with $r_h\lesssim0.5$ arcmin are recovered in approximately 90\% of the realizations, whereas the recovery probability decreases substantially for more extended populations. The absence of a significant spatial excess therefore places stronger constraints on compact, centrally concentrated systems than on extended or low-contrast populations. In particular, a spatial non-detection does not by itself exclude a more diffuse stellar system whose observable stars are distributed over a larger area than the adopted $2$-arcmin aperture. This interpretation is also robust to several variations of the injection experiment. Modest offsets between the adopted and true centres have relatively little effect on the recovery of compact populations, although offsets approaching $1$ arcmin substantially reduce their detectability. Experiments in which the injected population is divided between a compact component and a more extended circular or elongated component similarly show declining recovery as the fraction of stars in the compact component decreases. These tests are phenomenological rather than physical models of cluster disruption, but they demonstrate that the present analysis becomes progressively less sensitive as an observable population loses a compact central concentration.

The sensitivity of this test is limited by the depth, crowding, and completeness of the available VVVX--\emph{Gaia} catalogues, particularly for a low-luminosity population projected against the dense Sgr field. Deeper high angular-resolution imaging can provide a substantially stronger test by extending the analysis to fainter stellar populations and enabling more sensitive CMD and spatial comparisons with the local background. High-resolution HST observations of selected Minni fields provide a substantially deeper and more spatially resolved test of their stellar populations and structural properties; a dedicated analysis of these data will be presented by Garro et al. (in preparation). Previous photometric analyses similarly reached differing conclusions for individual candidates. \citet{piatti2021globular} independently examined four objects that overlap with our KMOS sample. Minni~342 was considered a possible Sgr globular cluster, although with weaker evidence than Minni~332, while Minni~341 was suggested as a possible younger open cluster; the physical reality of Minni~335 and Minni~344 could not be established. Our spectroscopy does not reproduce this ordering. Minni~342 has one of the lowest Sgr-compatible fractions ($f_{\rm Sgr}=0.52$) and the largest robust RV scatter in our sample ($19.8~{\rm km\,s^{-1}}$), whereas Minni~335 and Minni~344 each have $f_{\rm Sgr}=0.91$. The latter two, however, do not have RV distributions that are statistically narrower than expected from the Sgr comparison population. These differences are not necessarily contradictory, since the photometric and spectroscopic analyses probe different properties and are affected differently by the structured Sgr background. This distinction is also relevant to recent compilations of the Sgr cluster population. \citet{wang2026properties} included the 20 Minni systems identified in the VVVX studies in a sample of 31 Sgr-associated globular clusters, although RVs and proper motions were
not available for the Minni objects in that compilation. The present KMOS observations provide radial-velocity measurements for sixteen of these candidate fields. While the predominance of Sgr-compatible velocities supports their association with the Sgr environment, the absence of statistically compelling field-by-field kinematic or spatial signatures indicates that such compatibility should not by itself be interpreted as confirmation that each candidate constitutes a distinct globular cluster. The fact that the strongest indications arise in different diagnostics is itself informative. The narrowest RV distributions occur toward Minni~147 and Minni~148, whereas the strongest spatial excess occurs toward Minni~329, and the candidates previously favoured from photometric analyses do not necessarily stand out spectroscopically. The joint $V_{\rm los}$--$[\mathrm{Fe/H}]$ distributions provide a similar picture (Figure ~\ref{fig:field_rv_feh}). Several fields show apparently compact chemo-kinematic groupings, whereas others are difficult to distinguish visually from the Sgr comparison population. Importantly, the fields showing the strongest apparent coherence in this plane are not necessarily those with the narrowest RV distributions. Thus, the different diagnostics do not identify a single, consistent subset of candidates as distinct from the surrounding Sgr population. A genuine cluster would ideally produce mutually consistent evidence in several dimensions, such as a localized spatial concentration together with a coherent velocity and chemical distribution. No candidate in the present sample shows such statistically compelling evidence across the available diagnostics. However, the injection--recovery experiments demonstrate that the spatial non-detections place substantially stronger constraints on compact, centrally concentrated systems than on extended or low-contrast populations. Deeper spectroscopic follow-up, combining higher spectral resolution with substantially larger stellar samples, will therefore be required to test for intrinsic kinematic and chemical coherence and to place stronger constraints on the physical nature of these candidates.
\section{Summary and conclusions}
\label{sec:conclusions}

We have presented a homogeneous KMOS spectroscopic follow-up of sixteen Minni stellar-cluster candidates projected against the central regions of the Sagittarius dwarf spheroidal galaxy. Our main results are summarized as follows:

\begin{itemize}

\item We obtained reliable RVs for all 341 KMOS targets observed toward the sixteen candidate fields, providing a homogeneous spectroscopic characterization of the full observed sample.
\item The spectroscopy reveals a prominent Sgr-compatible component across the candidate fields. In total, 256 of the 341 KMOS targets (75\%) lie within the independently defined Sgr velocity interval, $120<V_{\rm los}<170~{\rm km\,s^{-1}}$. The individual fields nevertheless show substantial variation in their RV
distributions and in the fraction of targets falling within this interval.
\item We derived empirical $[\mathrm{Fe/H}]$ estimates from the KMOS spectra using the \textsc{KROSS}-M full-spectrum model calibrated against APOGEE DR17. The resulting metallicities broadly overlap the Sgr field population and provide the first homogeneous spectroscopic metallicity characterization of the sixteen observed fields.

\item The KMOS metallicities are broadly consistent with the earlier photometric estimates of \citet{garro2021physical}: fourteen of the sixteen fields agree within their combined $2\sigma$ uncertainties, with a median difference of $[\mathrm{Fe/H}]_{\rm KMOS}-[\mathrm{Fe/H}]_{\rm phot}=-0.17$~dex.

\item The candidate fields show different degrees of apparent chemo-kinematic coherence in the joint $V_{\rm los}$--$[\mathrm{Fe/H}]$ plane. Minni~147 and Minni~148 have the most unusual RV widths relative to random samples from the Sgr comparison population, but their nominal probabilities ($p_{\rm narrow}\simeq0.02$) are not significant once the sixteen fields examined are taken into account.

\item An independent VVVX--\emph{Gaia} spatial comparison finds no statistically compelling localized stellar excess among the tested candidate positions. Minni~329 shows the strongest nominal spatial signal, but this does not remain significant after accounting for the 16 positions tested and is not accompanied by significant central concentration.

\item Taken together, the KMOS observations establish that the photometrically preselected targets predominantly trace the Sgr chemo-kinematic population, but the present spectroscopic and spatial evidence is insufficient to establish any individual candidate as a distinct stellar cluster. The strongest indications arise in different fields and from different diagnostics, with no candidate showing consistent evidence across the available measurements. Establishing the physical nature of these candidates will therefore require deeper, spatially complete imaging to resolve their stellar populations and larger spectroscopic samples to test for kinematic and chemical coherence against the structured Sgr background.
\end{itemize}
\section*{Acknowledgements}
We thank the anonymous referee for their constructive comments and suggestions, which have improved the clarity and quality of this manuscript. This work was funded by the Postdoctoral Talent Attraction Competition for Research Centers and Institutes of the Universidad Andrés Bello (UNAB) 2025, project Nº. DI-07-25/ATP.  M.G. gratefully acknowledges support from Fondecyt through grant 1240755. J.G.F-T gratefully acknowledges the support provided by ANID Fondecyt Regular No. 1260371. D.M. gratefully acknowledges support from the Center for Astrophysics and Associated Technologies CATA by the ANID BASAL projects ACE210002 and FB210003, by Fondecyt Project No. 1220724. J.B. and R.K. acknowledge support from FONDECYT Regular Projects Nos. 1240249 and 1261142.  C.C. acknowledges support by ANID BASAL project FB210003. I.P. acknowledges support from ANID BECAS/DOCTORADO NACIONAL 21230761. BD acknowledges support from ANID Basal project FB210003. This work is supported by the China-Chile Joint Research Fund (CCJRF No.2301) and the Chinese Academy of Sciences South America Center for Astronomy (CASSACA) Key Research Project E52H540301. ZG is supported by FONDECYT Iniciación project 11260176.  ZG and CM are funded by the project ALMA-ANID 31240014. DBS acknowledges support from the Programa de doctorado en Astrofísica y Astroinformática of Universidad de Antofagasta. BD acknowledges support from ANID Basal project FB210003 and also that this work was funded by the 2026 Regular Bridge Fund of the Universidad Andrés Bello (UNAB), project No. DI-01-26/FPFR.

\section*{Data Availability}
The radial-velocity catalog for all measured stars, including their membership classifications, is available in electronic form at the CDS via anonymous ftp.

\bibliographystyle{aa}
\bibliography{Ref}

\clearpage
\begin{appendix}

\section{Radial-velocity distributions of the candidate fields}
\label{app:rv_distributions} 
\begin{figure}
    \centering
    \includegraphics[width=1\linewidth]{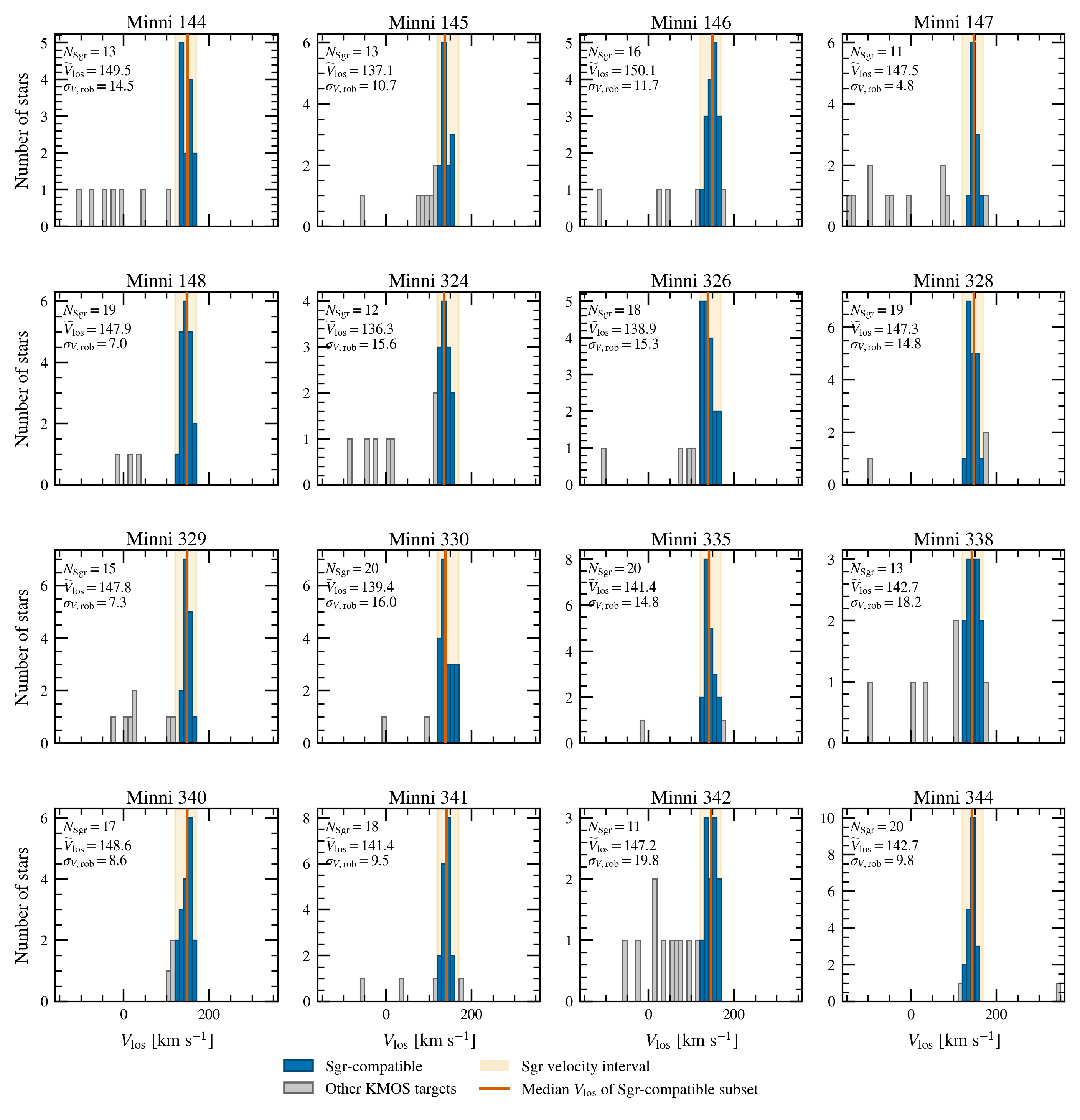}
    \caption{Line-of-sight velocity distributions of the KMOS targets toward the sixteen Minni candidate fields. Blue histograms show stars satisfying the fixed Sgr-compatible velocity criterion, $120<V_{\rm los}<170~{\rm km\,s^{-1}}$, while grey histograms show the remaining KMOS targets. The shaded region marks the adopted Sgr velocity interval, and the orange vertical line indicates the median velocity of the Sgr-compatible subset in each field. The annotations give the
number of Sgr-compatible stars, $N_{\rm Sgr}$, their median line-of-sight velocity, ${V}_{\rm los}$, and their robust velocity scatter, $\sigma_{V,\rm rob}$, defined as 1.4826 times the median absolute deviation. Most fields show a prominent concentration of targets within the Sgr-compatible velocity range, although the widths of these distributions vary between fields. The classification identifies stars kinematically compatible with the dominant Sgr population and does not imply membership in distinct stellar clusters.}
    \label{fig:rv_app1}
\end{figure}

Figures~\ref{fig:rv_app1} show the individual radial-velocity distributions of the sixteen candidate fields. Stars within the fixed interval $120<V_{\rm los}<170~{\rm km\,s^{-1}}$ are shown separately from the remaining KMOS targets. These distributions illustrate the field-level concentrations discussed in Sect.~\ref{sec:field_results}; the Sgr-compatible stars are not interpreted as confirmed cluster members.

\end{appendix}

\end{document}